\documentclass[manuscript,screen]{acmart}

\AtBeginDocument{%
  \providecommand\BibTeX{{%
    \normalfont B\kern-0.5em{\scshape i\kern-0.25em b}\kern-0.8em\TeX}}}

\newcommand{\mesh}{\Omega}
\newcommand{\timeint}{(t_0,t_f]}
\usepackage{algorithm}
\usepackage[noend]{algpseudocode}
\usepackage{graphicx,subcaption}
\usepackage[utf8]{inputenc}

\usepackage{tikz}
\usetikzlibrary{arrows,decorations.pathreplacing}
\usetikzlibrary{arrows.meta}
\usetikzlibrary{fit,positioning}

\DeclareFixedFont{\ttb}{T1}{txtt}{bx}{n}{12} 
\DeclareFixedFont{\ttm}{T1}{txtt}{m}{n}{12}  

\usepackage{xcolor}
\definecolor{maroon}{cmyk}{0, 0.87, 0.68, 0.32}
\definecolor{halfgray}{gray}{0.55}
\definecolor{ipython_frame}{RGB}{207, 207, 207}
\definecolor{ipython_bg}{RGB}{247, 247, 247}
\definecolor{ipython_red}{RGB}{186, 33, 33}
\definecolor{ipython_green}{RGB}{0, 128, 0}
\definecolor{ipython_cyan}{RGB}{64, 128, 128}
\definecolor{ipython_purple}{RGB}{170, 34, 255}

\usepackage{listings}
\lstdefinelanguage{Python}{
    morekeywords={access,and,break,class,continue,def,del,elif,else,except,exec,finally,for,from,global,if,import,in,is,lambda,not,or,pass,print,raise,return,try,while},%
    morekeywords=[2]{abs,all,any,basestring,bin,bool,bytearray,callable,chr,classmethod,cmp,compile,complex,delattr,dict,dir,divmod,enumerate,eval,execfile,file,filter,float,format,frozenset,getattr,globals,hasattr,hash,help,hex,id,input,int,isinstance,issubclass,iter,len,list,locals,long,map,max,memoryview,min,next,object,oct,open,ord,pow,property,range,raw_input,reduce,reload,repr,reversed,round,set,setattr,slice,sorted,staticmethod,str,sum,super,tuple,type,unichr,unicode,vars,xrange,zip,apply,buffer,coerce,intern},%
    sensitive=true,%
    morecomment=[l]\#,%
    morestring=[b]',%
    morestring=[b]",%
    morestring=[s]{'''}{'''},
    morestring=[s]{"""}{"""},
    morestring=[s]{r'}{'},
    morestring=[s]{r"}{"},%
    morestring=[s]{r'''}{'''},%
    morestring=[s]{r"""}{"""},%
    identifierstyle=\color{black}\ttfamily,
    commentstyle=\color{ipython_cyan}\ttfamily,
    stringstyle=\color{ipython_red}\ttfamily,
    keepspaces=true,
    showspaces=false,
    showstringspaces=false,
    frame=single,
    numbers=left,
    numberstyle=\tiny\color{halfgray},
    backgroundcolor=\color{ipython_bg},
    basicstyle=\scriptsize,
    keywordstyle=\color{ipython_green}\ttfamily,
    captionpos=b,
}

\definecolor{myorange}{rgb}{0.95,0.50,0.00} 

\definecolor{applegreen}{rgb}{0.55, 0.71, 0.0}

\setcopyright{acmcopyright}
\copyrightyear{2020}
\acmYear{2020}
\acmDOI{10.1145/1122445.1122456}

\begin{document}

\title{PyMGRIT: A Python Package for the parallel-in-time method MGRIT}

\author{Jens Hahne}
\email{jens.hahne@math.uni-wuppertal.de}
\orcid{1234-5678-9012}
\author{Stephanie Friedhoff}
\email{friedhoff@math.uni-wuppertal.de}
\author{Matthias Bolten}
\email{bolten@math.uni-wuppertal.de}
\affiliation{%
  \institution{Bergische Universit\"{a}t Wuppertal}
  \city{Wuppertal}
  \postcode{42097}
  \country{Germany}
}

\renewcommand{\shortauthors}{Hahne and Friedhoff, et al.}

\begin{abstract}
In this paper, we introduce the Python framework \texttt{PyMGRIT}, which implements the multigrid-reduction-in-time (MGRIT) algorithm for solving the (non-)linear systems arising from the discretization of time-dependent problems. The MGRIT algorithm is a reduction-based iterative method that allows parallel-in-time simulations, i.\,e., calculating multiple time steps simultaneously in a simulation, by using a time-grid hierarchy. The \texttt{PyMGRIT} framework features many different variants of the MGRIT algorithm, ranging from different multigrid cycle types and relaxation schemes, as well as various coarsening strategies, including time-only and space-time coarsening, to using different time integrators on different levels in the multigrid hierachy. The comprehensive documentation with tutorials and many examples, the fully documented code, and a large number of pre-implemented problems allow an easy start into the work with the package. The functionality of the code is ensured by automated serial and parallel tests using continuous integration. \texttt{PyMGRIT} allows serial runs for prototyping and testing of new approaches, as well as parallel runs using the Message Passing Interface (MPI). Here, we describe the implementation of the MGRIT algorithm in \texttt{PyMGRIT} and present the usage from both user and developer point of views. Three examples illustrate different aspects of the package, including pure time parallelism as well as space-time parallelism by coupling \texttt{PyMGRIT} with \texttt{PETSc} or \texttt{Firedrake}, which enable spatial parallelism through MPI.
\end{abstract}

\ccsdesc[500]{Mathematics of computing~Solvers}
\ccsdesc[300]{Mathematics of computing~Ordinary differential equations}
\ccsdesc[300]{Mathematics of computing~Partial differential equations}
\ccsdesc[100]{Mathematics of computing~Differential algebraic equations}

\keywords{Multigrid-reduction-in-time (MGRIT), parallel-in-time integration}

\maketitle

\section{Introduction}
\label{sec:intro}

The classical approach for solving an initial value problem is based on a time-stepping procedure, which computes the solution at a point in time based on the solution at one or more previous time steps and, thus, propagates the solution over time. The method is optimal, i.\,e., of order $\mathcal{O}\left(N_t\right)$ for $N_t$ time steps, and gives the solution after $N_t$ applications of the time integrator. However, time-stepping algorithms are completely sequential in time and limit potential parallelization to the spatial domain. In recent years, the structure of modern computer systems has been characterized by a growing number of processors, as the clock rates of the individual processors stagnate. As a consequence, possibilities of spatial parallelization are quickly exhausted, although further resources are available. Promising approaches for using these modern computer systems to further reduce the runtime of simulations of initial value problems are parallel-in-time methods, which allow not only spatial but also temporal parallelism.

One of these approaches is the iterative multigrid-reduction-in-time (MGRIT) algorithm \cite{RDFalgout_etal_2014}, which applies multigrid reduction principles to the time domain. The idea of the algorithm is based on using a hierarchy of time grids, where the finest grid contains the same points in time as in the time-stepping approach and the number of points on the coarser grids decreases, so that the coarsest grid consists of only a few points. While time integration is applied to the coarsest grid sequentially, time subdomains are handled in parallel on the other grids. Over the recent years, the algorithm has been successfully applied to various problems, e.\,g., linear and nonlinear parabolic problems \cite{RDFalgout_etal_2014,MR3716560}, compressible fluid dynamics  \cite{Falgout_2014_FAS}, power grids \cite{Lecouvez2016_powersystem, Schroder2018_powersystems}, eddy-current simulations \cite{friedhoff2019multigridreductionintime,bolten2019parallelintime,Friedhoff_2020aa}, linear advection \cite{HowseEtAl2019,sterck2019optimizing}, and machine learning \cite{doi:10.1137/19M1247620}.

Besides the MGRIT algorithm, there are many other iterative and direct parallel-in-time methods. Probably the most well-known method is Parareal \cite{JLLions_etal_2001a}, whose success is due to its simplicity and applicability: For the use of the algorithm only an expensive and a cheap time integrator is needed. This idea of Parareal can be interpreted in a variety of frameworks of numerical schemes. In particular, Parareal can be seen as a two-level MGRIT variant \cite{RDFalgout_etal_2014}.
The parallel full aproximation scheme in space and time (PFASST) \cite{emmett2012} is based on spectral deferred corrections (SDC) \cite{sdc} and allows space-time parallelism by simultaneously executing several SDC ``sweeps'' on a space-time hierarchy. Another method, the revisionist integral deferred correction method (RIDC) \cite{doi:10.1137/09075740X,10.1145/2964377} allows time parallelism through the pipelined parallel application of integral deferred corrections. A more detailed, structured, and general overview of parallel-time-integration approaches is given in \cite{Gander2015_Review}. Furthermore, the website \texttt{https://parallel-in-time.org} offers many more references and information about parallel-in-time methods.

Despite the increasing number of algorithms, concepts, and papers in the field of parallel-in-time integration, the number of accessible stand-alone parallel-in-time libraries providing parallelization in time or allowing space-time parallelism is relatively small. The following libraries are most noteable: The XBraid \cite{xbraid} library, written in C provides an implementation of the MGRIT algorithm and libridc \cite{10.1145/2964377} is a C\texttt{++} implementation of the RIDC algorithm. The PFASST method is implemented by various libraries which are summarized in \cite{pfasst}, in particular by the Fortran library \texttt{libpfasst}, \texttt{PFASST++} written in C\texttt{++}, and the Python implementation \texttt{pySDC}. Besides these, there are a number of other implementations of parallel-in-time concepts, but most of them are either more specialized or more difficult to access, such as the \texttt{SWEET} code \cite{sweet} or various implementations of the PARAEXP method \cite{doi:10.1137/110856137}. This paper describes the new Python framework \texttt{PyMGRIT}, which implements the MGRIT algorithm. 
\texttt{PyMGRIT} is designed to be easy to use, ranging from the very simple installation of the package to calling or overriding functions, making it ideal for testing the application of MGRIT to problems, or for prototyping new ideas and strategies for the MGRIT algorithm. More precisely, \texttt{PyMGRIT} allows on the one hand the simple, pure application of different variations of the MGRIT algorithm to problems without having to worry about implementation details, parallel communication, etc., but on the other hand, it allows flexible adaptations of individual components of the algorithm by overriding existing functions. This allows users to try MGRIT for their application, and it makes \texttt{PyMGRIT} particularly attractive for the training of students. Furthermore, \texttt{PyMGRIT} allows easy coupling with other software that already exists as Python packages, e.\,g., \texttt{Matplotlib} \cite{Matplotlib} for creating static and interactive visualizations or \texttt{Firedrake} \cite{Firedrake} for using the Finite Element Method (FEM). Additionally, \texttt{PyMGRIT} allows space-time parallel runs that go far beyond prototyping.

The following sections of this paper are organized as follows: Section \ref{sec:mgrit} describes the MGRIT algorithm. Based on this, Section \ref{sec:framework} introduces the \texttt{PyMGRIT} framework and describes the implementation of the algorithm and its variations. Then, the use of \texttt{PyMGRIT} is examined, describing first basic usage of the package, followed by a description of implementing a custom application using \texttt{PyMGRIT}. In Section \ref{sec:numerical_experiments}, different aspects and possibilities of using \texttt{PyMGRIT} are demonstrated in three numerical examples. Finally, conclusions are drawn and possible extensions of the package are discussed in Section \ref{sec:conclusion_and_outlook}.

\section{MGRIT}
\label{sec:mgrit}

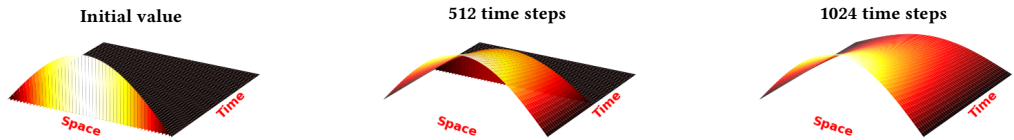
\begin{figure}
        \begin{tikzpicture}[y=1cm, x=1cm,ultra thick]    
			\node at (0,0){\includegraphics[width=.3\linewidth]{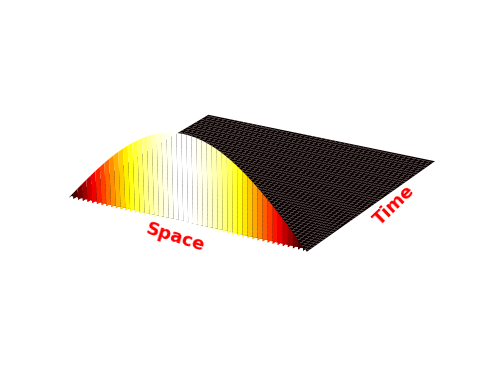}};
			\node at (5,0){\includegraphics[width=.3\linewidth]{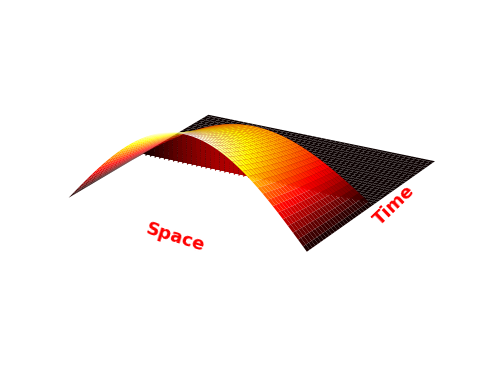}};
			\node at (10,0){\includegraphics[width=.3\linewidth]{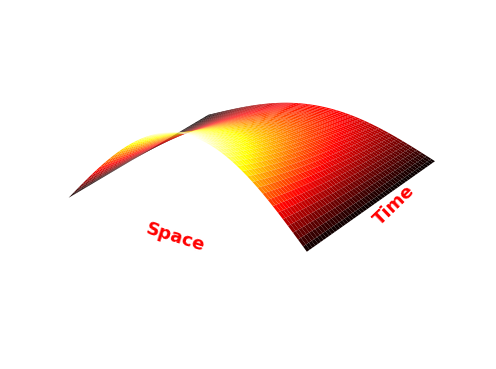}};
			\node[font=\footnotesize] at (0,1) {\textbf{Initial value}};
			\node[font=\footnotesize] at (5,1) {\textbf{512 time steps}};
			\node[font=\footnotesize] at (10,1) {\textbf{1024 time steps}};
			\end{tikzpicture}
	\caption{Sequential time stepping for a simple heat conduction problem. Starting with the initial condition (left), the solution is propagated over time (middle, right).}
	\label{fig:time_stepping}
\end{figure}

The idea of MGRIT is to enable parallelism in a traditionally sequential process. For time-dependent problems, starting with an initial value, the solution is classically computed step by step, i.\,e., starting with the initial condition, the solution at each time step is computed based on the solution at one or more previous time steps. This sequential time-stepping process is shown in Figure \ref{fig:time_stepping}, which displays the initial condition of a simple linear heat conduction problem along with the solutions after 512 and 1024 time steps. 

In contrast, the iterative MGRIT algorithm solves the problem by updating the solution at many points in time simultaneously. Thereby, the initial guess of the solution can be chosen arbitrarily for all points in time. The idea of the algorithm is based on a multilevel hierarchy for the problem. While the finest level contains the same points in time as in the time-stepping approach, on coarse levels only a subset of these points are considered. Optimally, the coarsest grid contains only a very small number of points in time, e.\,g., three or five, and can therefore be solved exactly and quickly by the classical time-marching approach. This solution on the coarsest grid is used in the algorithm to update the solution on the finer grids until the solution on the finest grid is accurate enough based on a predefined tolerance. More precisely, one iteration of the algorithm consists of the following steps:

\begin{enumerate}
\item Relaxation on the finest grid for multiple points in time simultaneously. Relaxation corresponds to a local application of the time integrator, which is also used for time stepping.
\item Transferring the solution to the next coarser grid, which contains only a subset of the points in time.
\item Repeating steps 1 and 2 for the next coarser grid until the coarsest grid is reached.
\item Solving the problem on the coarsest grid directly.
\item Interpolating the solution from the coarsest grid to the finest grid, improving the solution on all grid levels.
\end{enumerate}

\begin{figure}
	\resizebox{.99\linewidth}{!}{ %
        \begin{tikzpicture}[y=1cm, x=1cm,ultra thick]    
			\node (imgA){\includegraphics[width=2.75cm]{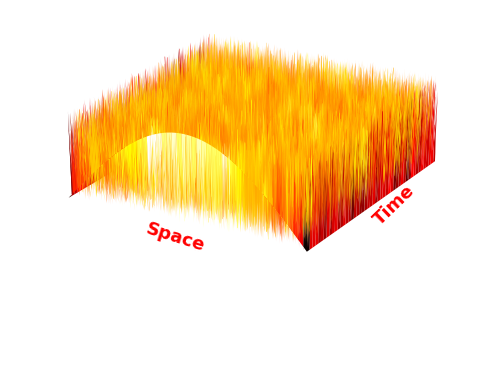}};
			\node[right=-0.5cm of imgA] (imgB){\includegraphics[width=2.75cm]{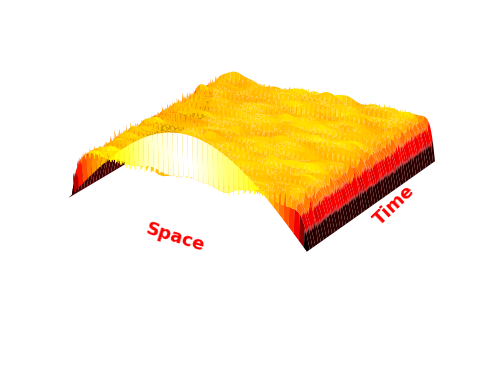}};
			
			\node[right=0.1mm of imgA]  (circleA) {};
			\node[left=0.1mm of imgB]  (circleB) {};
			\draw[-latex] (circleB) -- (circleA);			
			
			\node[right=0.5cm of imgB] (imgC){\includegraphics[width=2.75cm]{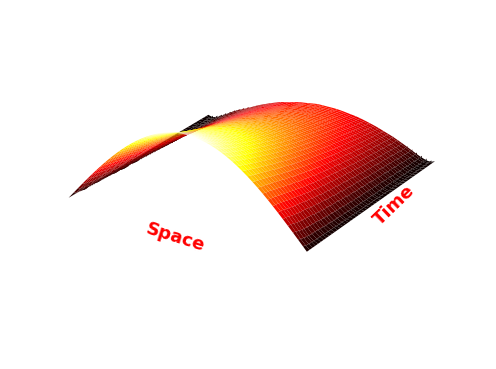}};

			\node at (1.375,1) {relaxation};
			
			\draw[draw=black] (-1.1,1.2) rectangle (3.75,-0.75);
			\draw[draw=black] (5,1.2) rectangle (7.25,-0.75);
			
			\draw[draw=black] (3.5,-1.75) rectangle (4,-2);
			\draw[draw=black] (4.25,-3) rectangle (4.75,-3.25);	
			\draw[draw=black] (4.75,-1.75) rectangle (5.25,-2);
			
			\draw[-Latex] (3.375,-0.75) to (3.875,-1.75);
			\draw[-Latex,dashed] (4,-2) to (4.5,-3);
			
			\draw[-Latex] (5.125,-1.75) to (5.625,-0.75);
			\draw[-Latex,dashed] (4.5,-3) to (5,-2);
			
			\node at (2.9,-1.25) {restriction};  
			\node at (6.5,-1.25) {interpolation};  
			
			\def \f {9.5}			
			
			\node[right=0.5cm of imgC] (imgA){\includegraphics[width=2.75cm]{mgrit_after_0.png}};
			\node[right=-0.5cm of imgA] (imgB){\includegraphics[width=2.75cm]{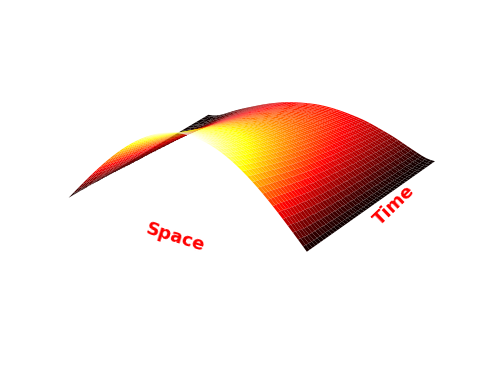}};
			
			\node[right=0.1mm of imgA]  (circleA) {};
			\node[left=0.1mm of imgB]  (circleB) {};
			\draw[-latex] (circleB) -- (circleA);			
			
			\node[right=0.5cm of imgB] (imgC){\includegraphics[width=2.75cm]{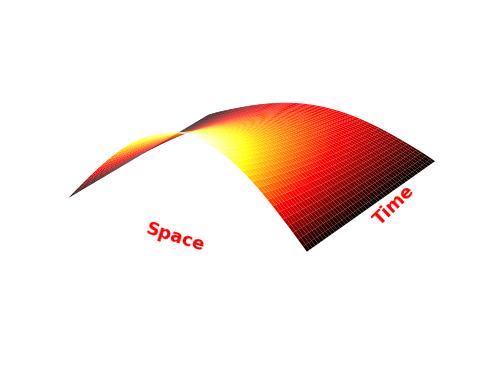}};

			\node at (\f+1.375,1) {relaxation};
			
			\draw[draw=black] (\f+-1.1,1.2) rectangle (\f+3.75,-0.75);
			\draw[draw=black] (\f+5,1.2) rectangle (\f+7.25,-0.75);
			
			\draw[draw=black] (\f+3.5,-1.75) rectangle (\f+4,-2);
			\draw[draw=black] (\f+4.25,-3) rectangle (\f+4.75,-3.25);	
			\draw[draw=black] (\f+4.75,-1.75) rectangle (\f+5.25,-2);
			
			\draw[-Latex] (\f+3.375,-0.75) to (\f+3.875,-1.75);
			\draw[-Latex,dashed] (\f+4,-2) to (\f+4.5,-3);
			
			\draw[-Latex] (\f+5.125,-1.75) to (\f+5.625,-0.75);
			\draw[-Latex,dashed] (\f+4.5,-3) to (\f+5,-2);
			
			\node at (\f+2.9,-1.25) {restriction};  
			\node at (\f+6.5,-1.25) {interpolation};  	
			
			\node[font=\large] at (3.075,1.75) {\textbf{Iteration 0}};  
			\node[font=\large] at (\f+3.075,1.75) {\textbf{Iteration 1}}; 		

   		\end{tikzpicture}}
	\caption{\label{fig:mgrit_iteration} Two MGRIT iterations for a simple heat conduction problem.}
\end{figure}

These steps are applied iteratively until a desired quality of the solution is achieved. Figure \ref{fig:mgrit_iteration} shows this process for two MGRIT iterations. While this is only a rough description of the algorithm, it reveals one of the key features of MGRIT: its non-intrusiveness. The only requirement for the algorithm is a time-stepping procedure, which is also used in the time-marching approach and, thus, this procedure already exists for many problems and can be integrated easily into a parallel framework with MGRIT.

After the previous schematic representation of time stepping and the MGRIT algorithm, we now want to examine both approaches algorithmically. Therefore, we consider an initial value problem of the form
\begin{equation}\label{eq:problem_system}
    	\mathbf{u}'(t) = \mathbf{f}(t,\mathbf{u}(t)), \quad \mathbf{u}(t_0) = \mathbf{g}_0, \quad t \in (t_0,t_f],
\end{equation} 
which can, for example, be a system of ordinary differential equations after the spatial discretization of a space-time partial differential equation. We discretize the time interval on a grid with uniformly distributed points $t_i = i\Delta t, \; i=0,1,\dots,N_t$ with constant time step $\Delta t = (t_f-t_0)/N_t$ and let $\mathbf{u}_i \approx \mathbf{u}(t_i)$ for $i = 0,\dots,N_t$ with $\mathbf{u}_0 = \mathbf{u}(0)$. Note that non-uniform distributions are also possible, but are not considered here for reasons of simplicity. Further, let $\boldsymbol{\Phi}_{i}$ be a time integrator, propagating the solution $\mathbf{u}_{i-1}$ from a time point $t_{i-1}$ to time point $t_i$, including forcing from the right-hand side. Then, the one-step time integration method for the time-discrete initial value problem \eqref{eq:problem_system} can be written as
\[
		\mathbf{u}_i = \boldsymbol{\Phi}_{i} (\mathbf{u}_{i-1}), \quad i = 1,2,\dots,N_t,
\]
or, considering all time points at once, as the space-time system
\begin{equation}\label{eq:mgrit_nonlinear_system_matrix}
		\mathcal{A}(\mathbf{u})\equiv
			\begin{bmatrix}
			\mathbf{u}_0\\
			\mathbf{u}_1-\boldsymbol{\Phi}_{1}(u_0)\\
			\vdots\\
			\mathbf{u}_{N_t} - \boldsymbol{\Phi}_{N_t}(u_{N_{t-1}})
		\end{bmatrix} = \begin{bmatrix}
			\mathbf{g}_0\\
			\mathbf{0}\\
			\vdots\\
			\mathbf{0}
		\end{bmatrix}\equiv \mathbf{g}.
\end{equation}
Sequential time-stepping solves problem \eqref{eq:mgrit_nonlinear_system_matrix} through a sequential forward solve, yielding the discrete solution after $N_t$ applications of the time integrator and, thus, sequential time-stepping is optimal, i.\,e., of order $\mathcal{O}\left(N_t\right)$. 

\begin{figure}
\begin{center}
	\begin{tikzpicture}
	
	\node at (-1.25,1.3) {Fine level}; 

	\draw[line width=1.2pt] (0,1.3) -- (8,1.3);
	\foreach \i in {0,.5,1,...,8}{
		\draw[line width=1.15pt] (\i,.125+1.3) -- (\i,-.125+1.3);
	}
	\foreach \i in {0,2,...,8}{
		\draw[line width=1.5pt] (\i,.225+1.3) -- (\i,-.225+1.3);
	}
	\foreach \i/\n in {0/0,.5/1,1/2}{
		\draw (\i,-.5+1.3) node {$t_\n$};
	}
	\draw (1.5,-.5+1.3) node {$\cdots$};
	\draw (2.1,-.5+1.3) node {$t_m$};
	\draw (8,-.5+1.3) node {$t_{N_t}$};
		
	\draw[line width=1.25pt,decorate,decoration={brace,amplitude=3pt},yshift=5pt] (5,-.5+1.3) -- (4.5,-.5+1.3) node[below=3pt,midway] {$\Delta t$};
	
	\node at (-1.25,0) {Coarse level}; 
	
	\draw[line width=1.2pt] (0,0) -- (8,0);
	\foreach \i in {0,2,...,8}{
		\draw[line width=1.15pt] (\i,.125) -- (\i,-.125);
	}
	\foreach \i in {0,2,...,8}{
		\draw[line width=1.5pt] (\i,.225) -- (\i,-.225);
	}

	\draw (0,-.5) node {$T_0$};
	\draw (2,-.5) node {$T_1$};
	\draw (3,-.5) node {$\cdots$};
	\draw (8,-.5) node {$T_{N_T}$};
			
	\draw[line width=1.5pt,decorate,decoration={brace,amplitude=5pt},yshift=5pt] (6,-.5) -- (4,-.5) node[below=7pt,midway] {$\Delta T = m\Delta t$};

	\end{tikzpicture}
	\caption{\label{fig:time_grid} Fine and coarse temporal grid with uniformly distributed points. The coarse grid is created by removing all $F$-points (represented by short markers) of the fine grid.}
	\end{center}
\end{figure}
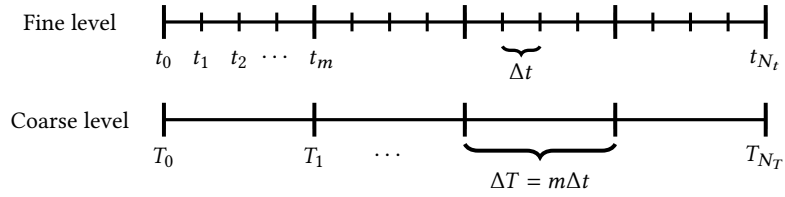

In order to embed the time integrator $\boldsymbol{\Phi}_{i}$ into a multi-level algorithm that iteratively solves problem \eqref{eq:mgrit_nonlinear_system_matrix}, we first define the components of the algorithm and then construct the algorithm. More precisely, we define a coarse grid system, a restriction and a prolongation operator between temporal grids, and a relaxation scheme. For a given fine temporal grid $t_i = i\Delta t, \; i=0,1,\dots,N_t$, and a given integer coarsening factor $m>1$, we define a splitting of the fine grid points into $F$- and $C$-points, where every $m$-th point is a $C$-point and all other points are $F$-points. Looking only at the $C$-points, we obtain a coarser grid $T_{i_c} = i_c\Delta T, \; i_c=0,1,\dots,N_T$, with $N_T = N_t/m$ and time step $\Delta T = m \Delta t$, as shown in Figure \ref{fig:time_grid}.

\begin{figure}
\begin{center}
	\begin{tikzpicture}
	\draw[rounded corners,fill=blue!60,blue!60] (0.25, 0.175) rectangle (1.75, -0.175) {};
	\draw[rounded corners,fill=blue!60,blue!60] (2.25, 0.175) rectangle (3.75, -0.175) {};
	\draw[rounded corners,fill=blue!60,blue!60] (4.25, 0.175) rectangle (5.75, -0.175) {};
	\draw[line width=1.2pt, -, >=latex'](0,0) -- coordinate (x axis) (6,0) node[right] {}; 

	\foreach \x in {0,1,...,12} 
		\draw[ultra thick, black] (0.5*\x,0.125) -- (0.5*\x,-0.125) node[below] {$t_{\x}$} ;
	
	\foreach \x in {0,4,8,12} 
		\draw[ultra thick, black] (0.5*\x,0.225) -- (0.5*\x,-0.225) node[below] {} ;

	\foreach \x in {1,2,3,5,6,7,9,10,11}{
		\node (a) at (-0.25+\x*0.5,0.6) {$\boldsymbol{\Phi}_{\x}$};
	}

	\foreach \x in {0,1,2}{
		\foreach \y in {0,1,2}{
			\node (point_00) at (\x*2-0.1+\y*0.5,0.2) {};
			\node (point_01) at (\x*2+0.6+\y*0.5,0.2) {};
			\draw[-{>[scale=0.7]}, ultra thick] (point_00) to[bend left = 40] (point_01);
	}}


	\def \s {8.0}

	\draw[rounded corners,fill=blue!60,blue!60] (9.85, 0.225) rectangle (10.15, -0.225) {};
	\draw[rounded corners,fill=blue!60,blue!60] (11.85, 0.225) rectangle (12.15, -0.225) {};
	\draw[rounded corners,fill=blue!60,blue!60] (13.85, 0.225) rectangle (14.15, -0.225) {};
	\draw[line width=1.2pt, -, >=latex'](\s,0) -- coordinate (x axis) (\s+6,0) node[right] {}; 
	\foreach \x in {0,1,2,3,4,5,6,7,8,9,10,11,12} 
		\draw[ultra thick, black] (\s+0.5*\x,0.125) -- (\s+0.5*\x,-0.125) node[below] {$t_{\x}$} ;
	\foreach \x in {0,4,8,12} 
		\draw[ultra thick, black] (\s+0.5*\x,0.225) -- (\s+0.5*\x,-0.225) node[below] {} ;

	\foreach \x in {4,8,12}{
		\node (a) at (\s-0.25+\x*0.5,0.6) {$\boldsymbol{\Phi}_{\x}$};
	}

	\foreach \y in {0,1,2}{
		\node (point_00) at (\s+1.4+2*\y,0.2) {};
		\node (point_01) at (\s+2.1+2*\y,0.2) {};
		\draw[-{>[scale=0.7]}, ultra thick] (point_00) to[bend left = 40] (point_01);
	}

	\node (a) at (3,-0.8) {$F$-relaxation};
	\node (a) at (\s+3,-0.8) {$C$-relaxation};

	\end{tikzpicture}
	\caption{\label{fig:relaxation} $F$- and $C$-relaxation for a temporal grid with 13 time points and a coarsening factor of four. In both relaxations, the independent intervals (blue) can be updated simultaneously.}
	\end{center}
\end{figure}
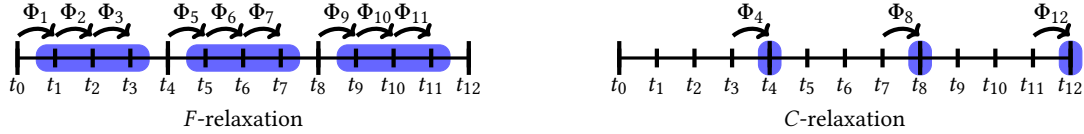

As the time-step size is different on the coarse grid, a time integrator $\boldsymbol{\Phi}_{i_c}$ is needed for the coarse level. We choose a re-discretization of the problem with time step $\Delta T$, but several choices are possible, such as coarsening in the order of the discretization~\cite{Nielsen2018,Falgout_etal_jcomp2019}. Next, we define two types of relaxation strategies, i.\,e., schemes of local applications of the time integrator. The first scheme, the so-called $F$-relaxation, performs a relaxation of all $F$-points by propagating the solution from a $C$-point to all following $F$-points up to the next $C$-point. Each interval of $F$-points can be executed in parallel, whereby each interval consists of $m-1$ sequential applications of the time integrator. The second scheme, the $C$-relaxation, analogously performs a relaxation of all $C$-points consisting of propagating the solution from the preceding $F$-point to a $C$-point. Again, all intervals of $C$-points can be updated simultaneously. Both relaxation schemes are shown in Figure \ref{fig:relaxation} and can be combined to new schemes. The $FCF$-relaxation, i.\,e., an $F$-relaxation followed by a $C$- and another $F$-relaxation, is the typical choice for the MGRIT algorithm, but other combinations are also possible. Finally, we define two operators for the transfer between the temporal fine and coarse grid, more precisely a restriction and a prolongation operator. While restriction is an injection, we define the ``ideal'' prolongation as an injection at $C$-points followed by an $F$-relaxation.

Using the full approximation storage (FAS) framework \cite{Brandt_1977_FAS} for solving both linear and nonlinear problems, the two-level MGRIT-FAS algorithm \cite{Falgout_2014_FAS} extended by spatial coarsening \cite{MR3716560} can be written as in Algorithm \ref{alg:mgrit}.

\begin{algorithm}
\caption{\label{alg:mgrit}MGRIT-FAS($\mathcal{A}, \mathbf{u}, \mathbf{g}$)}
\begin{algorithmic}[1]
\State Apply $F$-relaxation to $\mathcal{A}_1(\textbf{u}^{(1)})= \textbf{g}^{(1)}$
\State For $0$ to $\nu$:
\State \qquad Apply $CF$-relaxation to $\mathcal{A}_1(\textbf{u}^{(1)})= \textbf{g}^{(1)}$
\State Inject the approximation and its residual to the coarse grid: 
\Statex \qquad $\textbf{u}^{(2)} = \mathbf{R}_I(\textbf{u}^{(1)}), $
\Statex \qquad $\textbf{g}^{(2)} = \mathbf{R}_{I}(\textbf{g}^{(1)}-\mathcal{A}_1\textbf{u}^{(1)})$
\State If spatial coarsening:
\Statex \qquad $\textbf{u}^{(2)} = \mathbf{R}_s(\textbf{u}^{(2)})$ 
\Statex \qquad $\textbf{g}^{(2)} = \mathbf{R}_s(\textbf{g}^{(2)})$ 
\State Solve $\mathcal{A}_{2}(\textbf{v}^{(2)})= \mathcal{A}_{2}(\textbf{u}^{(2)}) + \textbf{g}^{(2)}$
\State Compute the error approximation: $\mathbf{e} = \mathbf{v}^{(2)} - \mathbf{u}^{(2)}$
\State If spatial coarsening:
\Statex \quad $\mathbf{e} = \mathbf{P}_s(e)$
\State Correct using ideal interpolation: $\textbf{u}^{(1)} = \textbf{u}^{(1)} + \mathbf{P}(\mathbf{e}) $
\end{algorithmic}
\end{algorithm}

In algorithm \ref{alg:mgrit}, $\mathcal{A}_l(\textbf{u}^{(l)})= \textbf{g}^{(l)}$ specifies the space-time problem on levels $l=1,2$, which can be either linear or nonlinear, and the two operators $\mathbf{R}_I$ and $\mathbf{P}$ denote the transfer operators between temporal grids. Additionally, the algorithm allows for spatial coarsening and we define $\mathbf{R}_s$ as spatial restriction and $\mathbf{P}_s$ as spatial prolongation. The relaxation scheme can be controlled by the parameter $\nu$, whereby at least one $F$-relaxation is performed per iteration and then $\nu$ subsequent $CF$-relaxations are executed. Typically $\nu = 1$, i.\,e., an $FCF$-relaxation scheme, is chosen for the MGRIT algorithm. Note that the two-level variant with $F$-relaxation is equivalent to the parareal method \cite{JLLions_etal_2001a}. To extend the two-level MGRIT-FAS algorithm to a multilevel setting, the two-level algorithm can be called resursively in step 6. Depending on the type and number of recursive calls, different multi-level variants can be defined. 

The two-level MGRIT-FAS algorithm is based on an initial guess. This initial guess can be chosen arbitrarily; however, a good initial guess offers natural advantages for the runtime convergence of the algorithm. If prior knowledge of the solution exists, it should be chosen as an initial guess. If nothing is known, an improved initial guess can be computed by the nested iteration \cite{GDahlquist_nes_it,LKronsjo_nes_it} strategy. The idea of nested iteration is to compute an initial approximation on the coarsest level and to interpolate it to the finer levels. Note that, independent of the initial guess, the MGRIT algorithm provides the exact discrete solution of the fine grid after $N_t/m$ or $N_t/\left(2m\right)$ iterations for $F$- or $FCF$-relaxation, respectively \cite{RDFalgout_etal_2014}.

\section{The PyMGRIT framework}
\label{sec:framework}

In the previous section, we have familiarized ourselves with the way MGRIT works and how the algorithm looks like. In this section, we introduce the \texttt{PyMGRIT} framework. First, we describe the structure and availability of the framework. We then present the implementation of the MGRIT algorithm in \texttt{PyMGRIT}. In terms of algorithmic parameters, many choices must be made such as the relaxation scheme or the cycling strategy that lead to many different variants of the algorithm. \texttt{PyMGRIT} pursues two goals: On the one hand, \texttt{PyMGRIT} should enable the use of MGRIT with a minimal choice of algorithmic parameters. On the other hand, \texttt{PyMGRIT} should be as flexible as possible and give users the possibility to adjust all settings with little effort. Section \ref{sec:mgrit_in_pymgrit} gives an overview of the most important MGRIT parameters and their implementation in \texttt{PyMGRIT}. Pursuing the first goal, most parameters have default values, but various choices are possible for pursuing the second goal. Section \ref{sec:application_of_pymgrit} presents a simple but typical code example for an application, while in Section \ref{sec:developer_view}, we take a developer's view of \texttt{PyMGRIT} and describe the classes required for implementing a custom application.

\subsection{Availability and structure}
\label{sec:availability_and_structure}

The \texttt{PyMGRIT} framework consists of three components: the code repository \cite{pymgrit-github}, the documentation \cite{pymgrit-docu}, and the Python Package Index (PyPI) version \cite{pymgrit-pypi}. The installation of the package is typically done using \texttt{pip}, and only a few requirements have to be fulfilled. First, a Python version $\geq 3.6$ is required and second, the following packages must be installed: \texttt{NumPy} \cite{Numpy}, \texttt{SciPy} \cite{2020SciPy-NMeth}, \texttt{Matplotlib} \cite{Matplotlib}, and \texttt{mpi4py} \cite{DALCIN20111124}. While \texttt{NumPy} and \texttt{SciPy} are used throughout the complete package, \texttt{Matplotlib} is mainly used for the visualization in different examples. The package \texttt{mpi4py} provides Python bindings for parallelization using the Message Passing Interface (MPI) standard. The requirements are listed in the file \texttt{setup.py} and will be installed automatically if PyPI and \texttt{pip} are used for the installation of the \texttt{PyMGRIT} package.

The code repository on GitHub \cite{pymgrit-github} contains the latest version of the current code. After each commit on the master branch, \texttt{GitHub Actions} are used as a continuous integration tool to trigger automated processes. Automated serial and parallel tests are performed for all Python versions $\geq 3.6$ using \texttt{pytest} \cite{pytest} and \texttt{tox} \cite{tox_website}. Additionally, the code coverage of the tests is calculated and the coverage is automatically uploaded to Codecov \cite{pymgrit-codecov}, where the results can be viewed. Furthermore, an up-to-date version of the documentation is created using Sphinx. The documentation consists of two parts, a pure application programming interface (API) documentation and a documentation of the features of \texttt{PyMGRIT}. While the API documentation is automatically generated from the Python comments, the feature documentation includes a quick start, tutorial, and many examples for both basic and advanced usage of \texttt{PyMGRIT}. By creating a new release on GitHub, the current repository version is automatically uploaded to PyPI and is available via \texttt{pip} afterwards.

The \texttt{PyMGRIT} package primarily consists of four directories:
\begin{itemize}
\item src: this directory contains the core features of \texttt{PyMGRIT}, namely implementations of the MGRIT algorithm and of template classes of other required components. It also contains a number of sample problems and their implementation.
\item examples: in this directory, a number of application examples can be found, both for basic and advanced usage of \texttt{PyMGRIT}, including all examples from the tutorial and from the documentation.
\item tests: the test directory contains all written tests that are automatically executed after each commit on the master branch. Tests include core functions of \texttt{PyMGRIT} as well as sample applications.
\item docs: contains the documentation for the examples, quick start, tutorial, etc.
\end{itemize}

\subsection{MGRIT in PyMGRIT}
\label{sec:mgrit_in_pymgrit}

\texttt{PyMGRIT} is based on classes, with the exception of some auxiliary functions. 
Two steps are required to use the MGRIT algorithm: First, an instance of the \texttt{Mgrit} class from \texttt{PyMGRIT}'s core must be created. In this step, all algorithmic choices and settings can be selected by parameters. The MGRIT implementation of \texttt{PyMGRIT} offers high flexibility for different variants of the algorithm, but at the same time the possibility to solve a problem with minimal specifications by using default values for most parameters. The second step is the actual solving of the problem, which is executed by simply calling a method of the \texttt{Mgrit} class.

Listing \ref{lst:mgrit_minimal} provides a minimal example of creating an instance of the class with subsequent solving of the problem. Only the problem hierachy has to be passed to the class via constructor parameters in line 3 (the structure of a problem hierarchy is described in section \ref{sec:problem_and_coarsening}). Other parameters of the constructor have default values, but can also be set manually. In line 4, the problem is then solved by calling the member function \texttt{solve}. The most important parameters of \texttt{PyMGRIT}'s core class \texttt{Mgrit} and their effects are presented below; a complete list of all parameters can be found in the documentation.

\begin{lstlisting}[float,language=Python,caption={Example for the minimal generation of an instance of the class Mgrit. Only a problem hierarchy is required, all other parameters have default values.}, label={lst:mgrit_minimal}]
from pymgrit import Mgrit

mgrit = Mgrit(problem=problem)
mgrit.solve()
\end{lstlisting}

\subsubsection{Problem hierarchy and coarsening}
\label{sec:problem_and_coarsening}

The only mandatory parameter for instantiating an object of the \texttt{Mgrit} class is a problem hierarchy. This problem hierarchy is a list of problems (objects of classes that inherit from \texttt{PyMGRIT}'s core \texttt{Application} class) that defines the time-multigrid hierarchy. The exact structure of the problems and how to implement a problem in an application class is described in Section \ref{sec:developer_view}. At this point, it is sufficient to know that the time-multigrid hierarchy, i.\,e., the number of levels and the coarsening factor for the MGRIT solver, is given by this problem hierarchy. The problem for each level contains a vector with all time points for that level. The only requirement for the time points of each coarse level is that the time points are a subset of the time points of the finest level. This allows great flexibility in the choice of the coarsening factor, as it allows the implementation of regular coarsening strategies, coarsening strategies with different coarsening factors per level, and also semi-coarsening strategies very easily. 

\subsubsection{Cycling and relaxation strategies}
\label{sec:cycling_and_relaxation_strategies}

\begin{figure}
\begin{center}
\begin{tikzpicture}

	\def \l {3.0}
	\foreach \x in {0,1,...,\l}{
		\fill (0.3*\x,0.5*\x) circle (3pt);
		\fill (-0.3*\x,0.5*\x) circle (3pt);
		\draw[-, line width=2pt]  (0,0) -- (0.3*\x,0.5*\x);
		\draw[-, line width=2pt]  (0,0) -- (-0.3*\x,0.5*\x);
	}

	\coordinate (last) at (2.1,1.5);
	\foreach \x/\y in {2.1/1.5, 2.4/1, 2.7/0.5, 3/0, 3.3/0.5, 3.6/0, 3.9/0.5, 4.2/1, 4.5/0.5, 4.8/0, 		5.1/0.5, 5.4/1, 5.7/1.5}{
		\fill (\x,\y) circle (3pt);
		\draw[-, line width=2pt]  (\x,\y) -- (last);
		\coordinate (last) at (\x,\y);
	}	
	
\end{tikzpicture}
	\caption{\label{fig:v_f_cycle} Structure of $V$- and $F$-cycles for four grid levels.}
	\end{center}
\end{figure}
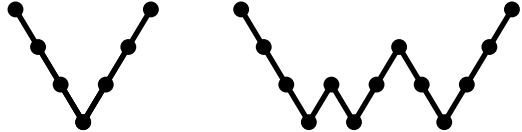

Two of the most important parameters for MGRIT are the cycling and the relaxation strategy. Both parameters have a significant impact on the convergence, parallelism, and efficiency of the algorithm. \texttt{PyMGRIT} provides two cycling strategies: $V$- and $F$-cycles. These can be controlled by the parameter \texttt{cycle\_type}, i.\,e., by adding the cycling strategy via \texttt{cycle\_type=`V'} or \texttt{cycle\_type=`F'} to the constructor call. The structure of the two cycles is shown in Figure \ref{fig:v_f_cycle}, which differs primarily in how often coarse grids are used. The $V$-cycle is the recursive implementation of algorithm \ref{alg:mgrit}, visiting the coarsest grid only once per MGRIT iteration. In contrast, the $F$-cycle visits the coarsest level several times per iteration, going back to the coarsest grid after having reached each level for the first time. As a consequence, an $F$-cycle better approximates the two-level method and, thus, faster convergence can be expected for $F$-cycles. However, $F$-cycles require more communication on intermediate coarse grids and more serial work on the coarsest grid, leading to worse parallel scalability compared to $V$-cycles for large numbers of processes.

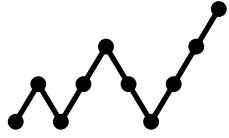
\begin{figure}
\begin{center}
\begin{tikzpicture}
	\coordinate (last) at (0,0);
	\foreach \x/\y in {0/0, 0.3/0.5, 0.6/0, 0.9/0.5, 1.2/1, 1.5/0.5, 1.8/0, 2.1/0.5, 2.4/1, 2.7/1.5}{
		\fill (\x,\y) circle (3pt);
		\draw[-, line width=2pt]  (\x,\y) -- (last);
		\coordinate (last) at (\x,\y);
	}	
\end{tikzpicture}
	\caption{\label{fig:nested_iterations} Structure of nested iteration when using four grid levels.}
	\end{center}
\end{figure}

Furthermore, an improved initial guess can be computed by using the nested iteration strategy, which can be controlled by the \texttt{Mgrit} constructor parameter \texttt{nested\_iteration}. By default, nested iteration is active, but it can be disabled by passing \texttt{nested\_iteration=False} to the constructor. As described at the end of Section \ref{sec:mgrit}, nested iteration computes an initial solution guess on the fine grid from coarse grids. More precisely, starting on the coarsest grid, an approximate solution is computed and interpolated to the next finer grid until the finest grid is reached. On each grid level, one $V$-cycle is used to compute an approximate solution, except on the coarsest grid, where the problem is solved directly. The resulting structure is illustrated in Figure \ref{fig:nested_iterations}. Note: In \texttt{PyMGRIT}, nested iteration is part of the setup phase of the algorithm, since it provides an initial solution guess on the finest grid.

The relaxation scheme in \texttt{PyMGRIT} can be set by the parameter \texttt{cf\_iter}. This parameter specifies how many $CF$-relaxations follow an always executed $F$-relaxation. By default, the parameter is set to $1$, i.\,e., an $FCF$-relaxation scheme. By varying this parameter, the user can easily switch between $F$-, $FCF$-, $FCFCF$-relaxation, and other relaxation schemes. Again, the choice of the scheme has a direct impact on the convergence of the algorithm. Usually, the more relaxation steps are performed, the better is the convergence. However, this improved convergence comes from the additional work performed in each iteration. The choice of the relaxation scheme also depends on the cycling strategy used. While an $FCF$-relaxation scheme is often a good starting point for a $V$-cycle, only $F$-relaxation is often sufficient for an $F$-cycle \cite{RDFalgout_etal_2014}. Note that the always performed $F$-relaxation on the finest grid is skipped after the second MGRIT iteration, because the updates during this $F$-relaxation are already performed during the post-relaxation of the ideal interpolation of the previous iteration.

\subsubsection{Stopping criterion}
\label{sec:stopping_criterion}

Another important parameter is the stopping tolerance of the algorithm, i.\,e., a threshold which stops the iterations if crossed. By default, \texttt{PyMGRIT} uses the absolute (space-)time residual of system \eqref{eq:mgrit_nonlinear_system_matrix} to check convergence. More precisely, denoting the residual of system \eqref{eq:mgrit_nonlinear_system_matrix} at the $i_c$-th $C$-point of the finest grid at iteration $k$ by $r^{(k)}_{i_c}$, where $i_c$ is an integer index for looping over all $C$-points, convergence is measured in the $2$-norm based on the absolute (space-)time residual,
\[
	\|r^{(k)}\| = \left(\sum_{i_c}\|r^{(k)}_{i_c}\|^2\right)^{1/2},
\]
where the computation of the norm of the residual at each $C$-point, $\|r^{(k)}_{i_c}\|$, is specified by the user in a class that inherits from \texttt{PyMGRIT}'s core \texttt{Vector} class (see Section \ref{sec:developer_view} for details). The convergence tolerance defines when to stop the iterations by checking $\|r^{(k)}\|  < \texttt{tol}$ at each iteration, where the tolerance can be set using the constructor parameter \texttt{tol}; by default, $\texttt{tol}$ is set to $10^{-7}$. Currently, the only stopping criterion implemented in \texttt{PyMGRIT} is based on the (space-)time residual; however, the documentation includes examples of how to customize and change this criterion.

\subsubsection{Parallel decomposition and computation}
\label{sec:parallel_decomposition_and_computation}

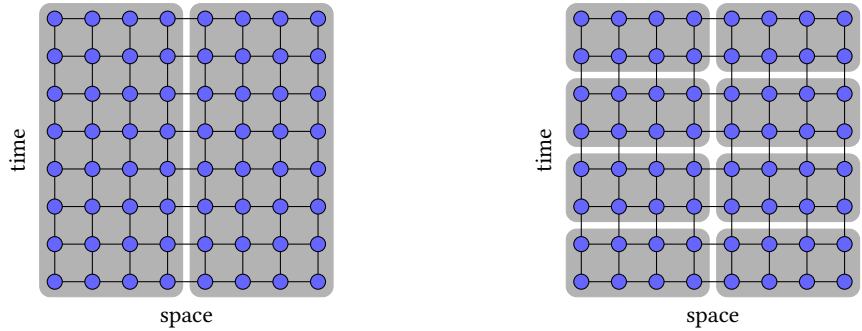
\begin{figure}
\begin{center}
\begin{tikzpicture}
\def \f {7.0}

\draw[rounded corners,fill=gray!60,gray!60] (-0.2, -0.2) rectangle (1.7, 3.7) {};
\draw[rounded corners,fill=gray!60,gray!60] (1.8, -0.2) rectangle (3.7, 3.7) {};

\foreach \a in {0, 1, ..., 3 }{
\draw[rounded corners,fill=gray!60,gray!60] (\f-0.2, \a-0.2) rectangle (\f+1.7, \a+0.7) {};
\draw[rounded corners,fill=gray!60,gray!60] (\f+1.8, \a-0.2) rectangle (\f+3.7, \a+0.7) {};
}

\foreach \a in {0, 0.5, ..., 3.5 }{
		\draw (\a,0) -- (\a,3.5);
		\draw (0,\a) -- (3.5,\a);
		\draw (\f+\a,0) -- (\f+\a,3.5);
		\draw (\f+0,\a) -- (\f+3.5,\a);}

\foreach \a in {0, 0.5, ..., 3.5 }{
	\foreach \b in {0, 0.5, ..., 3.5 }{
        \draw[black,fill=blue!60] ({\a}, {\b}) circle (0.1cm);
        \draw[black,fill=blue!60] ({\f+\a}, {\b}) circle (0.1cm);}}

\node at (1.75,-0.5) {space};
\node at (\f+1.75,-0.5) {space};

\node[rotate=90] at (-0.5,1.75) {time};
\node[rotate=90] at (\f-0.5,1.75) {time};

\end{tikzpicture}
	\caption{\label{fig:space_time_distribution} Decomposition of a space-time domain used for parallelizing computations in classical time-stepping algorithms (left) and in MGRIT (right).}
	\end{center}
\end{figure}

In implementations based on a classical time-stepping approach, typically solution values are stored for one time point (or for a few time points in case of a multistep method) at a time, and values are updated at each step of the method. Therefore, such algorithms only offer the possibility to decompose the computational domain of the problem in space and, thus, allow spatial parallelism only to speed up the calculation. On the contrary, \texttt{PyMGRIT} stores the solution at every point in time and, thus, \texttt{PyMGRIT} provides another dimension for the decomposition of the computational domain: the time dimension. Figure \ref{fig:space_time_distribution} shows both decompositions of a space-time domain, reflecting the parallelization strategies applied in classical space-parallel time-stepping algorithms and in space-time-parallel MGRIT. Notice, that the partitioning of the temporal domain into time slices is added to the spatial decomposition, allowing a distribution of the time slices across additional parallel processing units. \texttt{PyMGRIT} provides a function \texttt{split\_communicator} that splits a communicator into spatial and temporal components and returns two communicators for these components. The two communicators can be passed to the \texttt{Mgrit} constructor using the parameters \texttt{comm\_x} for the space communicator and \texttt{comm\_t} for the time communicator.

\begin{figure}
\begin{center}
	\begin{tikzpicture}
	
	\def \s {0.3}	
	\def \l {32.0}
	\def \b {-1.3}
	
	\draw[rounded corners,fill=blue!60,blue!60] (-\s*.33, -.25) rectangle (\l/2*\s+\s*.33, .25) {};
	\draw[rounded corners,fill=gray!60,gray!60] (\l/2*\s+\s-\s*.33, -.25) rectangle (\l*\s+\s*.33, .25) {};
	\node at (-1.25,0) {Level 0}; 

	\draw[line width=1.2pt] (0,0) -- (\l*\s,0);
	\foreach \i in {0,1,...,\l}{
		\draw[line width=1.15pt] (\s*\i,.125) -- (\s*\i,-.125);
	}
	\foreach \i in {0,4,...,\l}{
		\draw[line width=1.5pt] (\s*\i,.225) -- (\s*\i,-.225);
	}
	
	\draw[rounded corners,fill=blue!60,blue!60] (-\s*.33, -.25+\b) rectangle (\l/2*\s+\s*.33, .25+\b) {};
	\draw[rounded corners,fill=gray!60,gray!60] (\l/2*\s+\s-\s*.33, -.25+\b) rectangle (\l*\s+\s*.33, .25+\b) {};
	\node at (-1.25,\b) {Level 1}; 

	\draw[line width=1.2pt] (0,\b) -- (\l*\s,\b);
	\foreach \i in {0,4,...,\l}{
		\draw[line width=1.15pt] (\s*\i,.125+\b) -- (\s*\i,-.125+\b);
	}
	\foreach \i in {0,16,...,\l}{
		\draw[line width=1.5pt] (\s*\i,.225+\b) -- (\s*\i,-.225+\b);
	}
	
	\draw[rounded corners,fill=blue!60,blue!60] (-\s*.33, -.25+2*\b) rectangle (\l*\s+\s*.33, .25+2*\b) {};
	\node at (-1.25,2*\b) {Level 2}; 

	\draw[line width=1.2pt] (0,2*\b) -- (\l*\s,2*\b);
	\foreach \i in {0,16,...,\l}{
		\draw[line width=1.5pt] (\s*\i,.225+2*\b) -- (\s*\i,-.225+2*\b);
	}

	\draw[rounded corners,fill=blue!60,blue!60] (\l*\s+0.6,-.95) rectangle (\l*\s+1.1,-1.15) {};
	\draw[rounded corners,fill=gray!60,gray!60] (\l*\s+0.6,-1.45) rectangle (\l*\s+1.1,-1.65) {};
	\node at (\l*\s+2,-1.05) {Process 0};
	\node at (\l*\s+2,-1.55) {Process 1};
	\end{tikzpicture}
	\caption{\label{fig:parallel_time_distribution} Example of the distribution of time points across two processes in \texttt{PyMGRIT}. The time points of the finest level are distributed evenly, and the coarser levels contain all $C$-points of their parent fine grids. On the coarsest grid, one process holds all time points.}
	\end{center}
\end{figure}
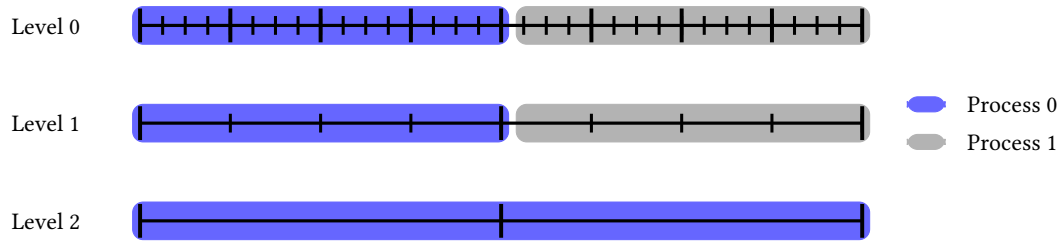

\texttt{PyMGRIT} distributes the time points of the finest grid equally over all processes contained in the time communicator. The distributions on coarser grids are based on the decompositions on their parent fine grids as each coarser grid contains a subset of their parent fine grid. The only exception is the coarsest grid, on which the problem is solved sequentially by one process. This process then distributes the solution values to other processes. Figure \ref{fig:parallel_time_distribution} shows the distribution of $N_t=33$ time points for two processes in a three-level setting with a coarsening factor of $m=4$.

Storing all time points in the MGRIT algorithm enables (additional) parallelism, but at the same time increases memory requirements. For each time point, \texttt{PyMGRIT} creates an object for the solution at this time point. Additionally, on coarser levels, two additional objects must be stored for each time point: one for a restricted copy of the fine-grid object and one for the FAS right-hand side. There are several ways to reduce the increased memory requirements for the MGRIT algorithm, e.\,g., storing only $C$-points. Yet, this is not implemented in \texttt{PyMGRIT}, but it is a possible extension for future versions.

\subsubsection{Spatial coarsening}
\label{sec:spatial_coarsening}

By default, \texttt{PyMGRIT} applies only temporal coarsening and does not require any additional information about the restriction or interpolation between spatial grids. If spatial coarsening is desired, the user can pass a list of spatial transfer operators using the \texttt{transfer} parameter. Thereby, a transfer operator is expected for each transfer between MGRIT levels, which ensures high flexibility for the user.

\subsubsection{Plots}
\label{sec:pymgrit_plots}

The \texttt{PyMGRIT} class \texttt{MgritWithPlots} provides an extension of the solver with several plotting functions, e.\,g., for visualizing MGRIT cycles or the distribution of time points across processes; example plots can be found in Section \ref{sec:example_1}.

\subsection{Basic usage of PyMGRIT}
\label{sec:application_of_pymgrit}

In this section, we describe how to solve a simple application problem and we explain the structure of a typical main routine of an application code that uses \texttt{PyMGRIT}. For a simple application problem, we choose the simplest case of a scalar ODE. More precisely, our goal is solving Dahlquist's test problem,
\begin{equation}\label{eq:dahlquist}
u' = \lambda u \;\;\text{ in } (0, 5]
\end{equation}
with $u(0) = 1$ and $\lambda = -1$. This and other application problems are included in \texttt{PyMGRIT}. For an overview of the applications included, see the online documentation. It provides a detailed tutorial and many examples for basic and advanced usage of \texttt{PyMGRIT}.

\begin{lstlisting}[float,language=Python,caption={Typical main routine of an application code that uses \texttt{PyMGRIT}.}, label={lst:main_routine}]
    # Create Dahlquist's test problem with 101 time steps in the interval [0, 5]
    dahlquist = Dahlquist(t_start=0, t_stop=5, nt=101)

    # Construct a two-level multigrid hierarchy for the test problem using a coarsening factor of 2
    dahlquist_multilevel_structure = simple_setup_problem(problem=dahlquist, level=2, coarsening=2)

    # Set up the MGRIT solver for the test problem and set the solver tolerance to 1e-10
    mgrit = Mgrit(problem=dahlquist_multilevel_structure, tol=1e-10)

    # Solve the test problem
    info = mgrit.solve()
\end{lstlisting}

The main routine of an application code that uses \texttt{PyMGRIT} to solve Dahlquist's test problem is presented in Listing \ref{lst:main_routine}. The program uses a two-level MGRIT algorithm to solve Dahlquist's test problem in the time interval $[0,5]$ with $101$ equidistant time points. The $101$ time points are composed of one point for the start time $t=0$ and 100 other time points. Note that one point for the start time is always included in the time interval in \texttt{PyMGRIT}. The structure of a main routine usually consists of three steps. First, the problem is created. Then, a multigrid hierarchy is constructed for this problem. Finally, the problem is solved using the MGRIT algorithm. In our example, for the first step, an instance of \texttt{PyMGRIT}'s class \texttt{Dahlquist} is created in line 2 that describes the fine problem. The time domain is passed to the problem class by using the parameters \texttt{t\_start} and \texttt{t\_stop} for specifying the time interval bounds and the parameter \texttt{nt} for the number of time steps. Afterwards, for the second step, a multilevel hierarchy is constructed in line 5, based on the problem instance \texttt{dahlquist}. Here, the auxiliary function \texttt{simple\_setup\_problem} is used and a two-level hierarchy with a coarsening factor of two is chosen. This results in a coarse level with 51 points in time. Note that \texttt{PyMGRIT} offers several ways to pass the time domain information to a problem class, as well as for creating a multilevel hierarchy; see the documentation for more details. For the third step, the MGRIT solver for the test problem is set up in line 8 as an instance of \texttt{PyMGRIT}'s core class \texttt{Mgrit} using the multilevel object \texttt{dahlquist\_multilevel\_structure}. Furthermore, the halting tolerance is set to $1e-10$. Finally, the problem is solved by calling the \texttt{solve} routine of the solver \texttt{mgrit} in line 11. The solver returns some statistical information about the run in the dictionary \texttt{info}.

\subsection{Developer view}
\label{sec:developer_view}

\subsubsection{Application structure}

\texttt{PyMGRIT} is based on four different types of classes:

\begin{itemize}
\item Solver: the solver class provides the implementation of the MGRIT algorithm. 
\item Vector: vector classes contain the solution of a single point in time. Every vector class must inherit from \texttt{PyMGRIT}'s core \texttt{Vector} class.
\item Application: application classes contain information about the problem we want to solve. Every application class must inherit from \texttt{PyMGRIT}'s core \texttt{Application} class.
\item GridTransfer: grid transfer classes contain information about the transfer of spatial grids between consecutive MGRIT levels. Every grid transfer class must inherit from \texttt{PyMGRIT}'s core \texttt{GridTransfer} class.
\end{itemize}

The three types of classes Vector, Application and GridTransfer are all based on abstract super classes. These classes independently create some structures that are valid for each type of class and also ensure that all necessary member variables and member functions exist in the respective child classes. A developer who wants to create and solve a problem that is not included in the \texttt{PyMGRIT} package usually only has to specify parts of the four classes. In most cases it is sufficient to write a vector class and an application class. The grid transfer class is primarily needed for the additional feature of spatial coarsening and, if spatial coarsening is not used, it is automatically created by the solver. For the most part, the solver class can be used without modifications, but changes are possible; see the documentation for examples.

\subsubsection{From time-stepping to PyMGRIT}

Listing \ref{lst:time_stepping} shows a typical time-stepping code for solving problem \ref{lst:main_routine} discretized by backward Euler on a temporal mesh with 101 points. The code consists of three components: first, the initial condition \texttt{value = 1} is set in line 2. The variable \texttt{value} is further used to store the propagated solution at the current time. The second component consists of the time information in lines 3 and 4. The temporal grid contains $nt = 101$ points in the time interval $[0,5]$ and is created using the \texttt{Numpy} function \texttt{linspace}. In the last step, we iterate over all points in time and apply the time integration in form of backward Euler. In summary, we have as components a variable that contains the solution at a point in time, time information belonging to the problem, and the time-integration loop.

\begin{lstlisting}[float,language=Python,caption={Example timestepping code for problem \ref{eq:dahlquist}, discretized by backward Euler on a temporal mesh with 101 points.}, label={lst:time_stepping}]
    constant_lambda = -1 # set lambda to -1
    value = 1  # initial solution value
    nt = 101  # number of time points
    t = np.linspace(0, 5, nt)  # time points: nt evenly spaced numbers in interval [0,5]
    # backward Euler time integration
    for i in range(1, nt): 
        value = 1 / (1 - (t[i] - t[i - 1]) * constant_lambda) * value
\end{lstlisting}

To implement these components in \texttt{PyMGRIT}, we first write a vector class \texttt{VectorDahlquist}, which stores the solution at a point in time and inherits from the \texttt{PyMGRIT} core class \texttt{Vector}. 
Therefore, we create a member variable \texttt{value}, which contains the solution of a point in time. Furthermore, the following member functions have to be implemented: \texttt{set\_values}, \texttt{get\_values}, \texttt{clone}, \texttt{clone\_zero}, \texttt{clone\_rand}, \texttt{\_\_add\_\_}, \texttt{\_\_sub\_\_}, \texttt{norm}, \texttt{pack}, and \texttt{unpack}. The function \texttt{set\_values} receives data values and overwrites the values of the vector data and \texttt{get\_values} returns the vector data. The function \texttt{clone} clones the object, \texttt{clone\_zero} returns a vector object initialized with zeros, and \texttt{clone\_rand} similarly returns a vector object initialized with random data. The functions \texttt{\_\_add\_\_}, \texttt{\_\_sub\_\_}, and \texttt{norm} define the addition and subtraction of two vector objects and the norm of a vector object, respectively.  Finally, the functions \texttt{pack} and \texttt{unpack} define the data to be communicated and how data is unpacked after receiving it. The definition of the class \texttt{VectorDahlquist} with implementations of the constructor and of the functions \texttt{\_\_add\_\_}, \texttt{\_\_sub\_\_} is shown in Listing \ref{lst:vector_class_dahlquist}. 
The implementation of the other functions is straightforward and not specified in detail here; please refer to the documentation for more information.

\begin{lstlisting}[float,language=Python,caption={Vector class for Dahlquist's test equation. Note: The definition of the class is not complete, the member functions \texttt{set\_values}, \texttt{get\_values}, \texttt{clone}, \texttt{clone\_zero}, \texttt{clone\_rand},  \texttt{norm}, \texttt{pack}, and \texttt{unpack} are not shown.}, label={lst:vector_class_dahlquist}]
class VectorDahlquist(Vector):
    def __init__(self, value):
        super().__init__()
        self.value = value
    
    def __add__(self, other):
    	return VectorDahlquist(self.get_values() + other.get_values())

    def __sub__(self, other):
        return VectorDahlquist(self.get_values() - other.get_values())
\end{lstlisting}

Second, we write an application class \texttt{Dahlquist} which contains information about the problem we want to solve. This class contains information about the time grid and the step function and is shown in Listing \ref{lst:application_class_dahlquist}. The time information is automatically provided by the \texttt{PyMGRIT} core class \texttt{Application}, from which every \texttt{PyMGRIT} application must inherit from; for more information see the tutorial. The function \texttt{step} must be defined and contains the time integration routine, which is the same as in Listing \ref{lst:time_stepping} except for names and accesses. To compute the new solution, the function receives as parameters the solution of the previous time point, \texttt{u\_start}, as well as the start point and the end point of the time integration step, \texttt{t\_start} and \texttt{t\_stop}, respectively. Furthermore, two mandatory member variables, \texttt{vector\_template} and \texttt{vector\_t\_start}, must be created in the application class. The variable \texttt{vector\_template} stores an instance of the corresponding Vector class, i.\,e., the \texttt{DahlquistVector} class, and \texttt{vector\_t\_start} defines the initial condition using the same class. This is all we need for our test problem. We can now use \texttt{PyMGRIT} to solve our problem as described in Listing \ref{lst:main_routine}. 

\begin{lstlisting}[float,language=Python,caption={Application class for Dahlquist's test equation.}, label={lst:application_class_dahlquist}]
class Dahlquist(Application):
    def __init__(self, constant_lambda = -1, *args, **kwargs):
        super().__init__(*args, **kwargs)
        self.constant_lambda = constant_lambda
        self.vector_template = VectorDahlquist(0) # Set the data structure for any time point
        self.vector_t_start = VectorDahlquist(1) # Set the initial condition

    # Time integration routine
    def step(self, u_start: VectorDahlquist, t_start: float, t_stop: float) -> VectorDahlquist:
        return VectorDahlquist(1 / (1 - (t_stop - t_start) * self.constant_lambda ) * u_start.get_values())
\end{lstlisting}

\section{Numerical experiments}
\label{sec:numerical_experiments}

In this section, we present several examples for using \texttt{PyMGRIT} to run simulations with MGRIT. Each example has its own focus and highlights different aspects of the framework. The first example gives an overview of runs with different variants of \texttt{PyMGRIT}'s MGRIT algorithm and shows some plots generated by \texttt{PyMGRIT}. The goal of this example is to provide a better understanding of the algorithm. The second example shows how \texttt{PyMGRIT} can dramatically reduce the simulation time of an existing realistic application. As an example, a two-dimensional electrical machine is used, where a single time step calls the \texttt{GetDP} \cite{PDular_etal_1998,getdp, getdp_website} solver. This example also uses spatial coarsening to further reduce the simulation time. The third example demonstrates the coupling of \texttt{PyMGRIT} and \texttt{PETSc} and shows space-time parallel results. The tests of all three examples were performed on an Intel Xeon Phi Cluster consisting of four 1.4 GHz Intel Xeon Phi processors.

\subsection{Using PyMGRIT for implementing various MGRIT variants}
\label{sec:example_1}

In the first example, we compare several variants of the MGRIT algorithm for the one-dimensional heat equation,
\[
u_t - au_{xx} = b(x,t) \;\; \text{ in } \; [0,1]\times[0,2],
\]
with thermal conductivity $a = 1$, right-hand side $b(x,t)=-\sin(\pi x) (\sin(t) - \pi^2 \cos(t))$, homogeneous Dirichlet boundary conditions in space, and subject to the initial condition $u(x,0) = \sin(\pi x)$.

The problem is discretized using second-order central finite differences with $1025$ degrees of freedom in space and on an equidistant time grid with $1024$ intervals using backward Euler. We choose five-level MGRIT algorithms with coarsening factor $m=4$ and consider the following variants:
\begin{enumerate}
\item $V$-cycles with $FCF$-relaxation,
\item $V$-cycles with $FCFCF$-relaxation,
\item $F$-cycles with $F$-relaxation,
\item $F$-cycles with $FCF$-relaxation, and
\item $F$-cycles with $FCFCF$-relaxation.
\end{enumerate}
For all variants, the stopping tolerance is set to $10^{-7}$ and a random initial guess is used. Note that we are not considering a five-level $V$-cycle with $F$-relaxation, since a multilevel setting with only $F$-relaxation generally does not provide an optimal algorithm \cite{RDFalgout_etal_2014}. The code for the example can be found in the \texttt{examples/toms} directory of \texttt{PyMGRIT}.

\begin{figure}
	\centerline{\includegraphics[width=.6\linewidth]{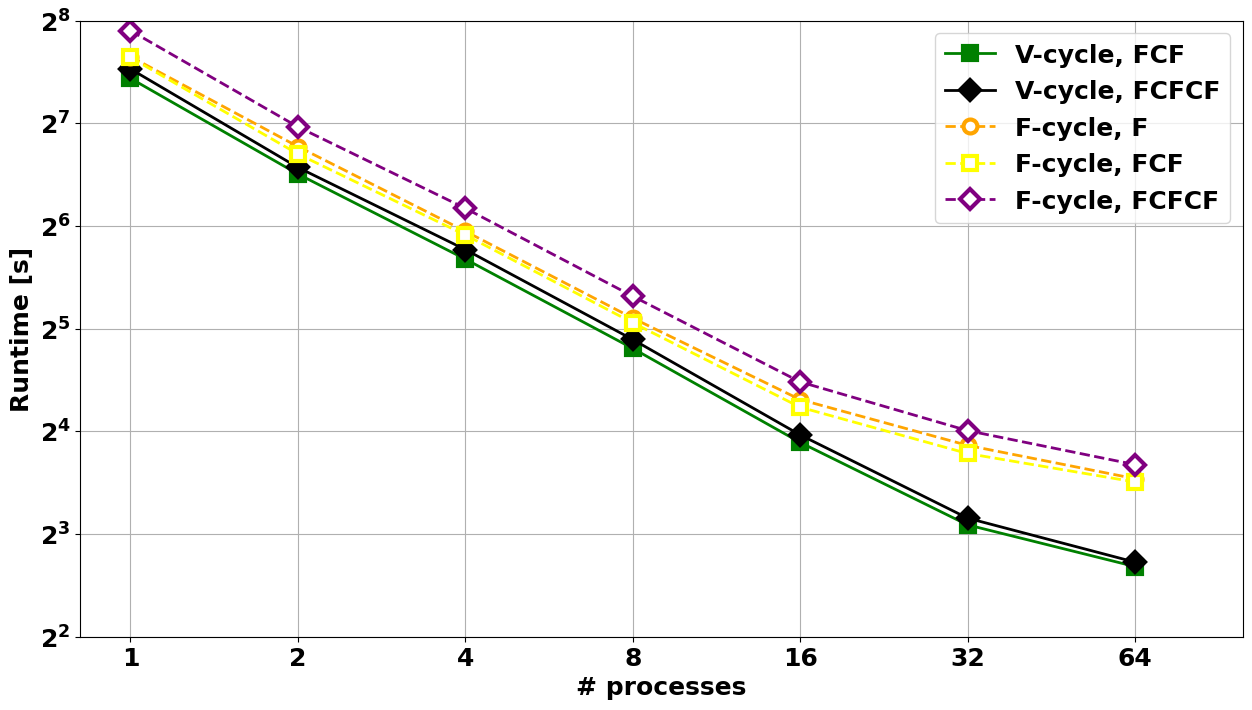}\quad
	\includegraphics[width=.39\linewidth]{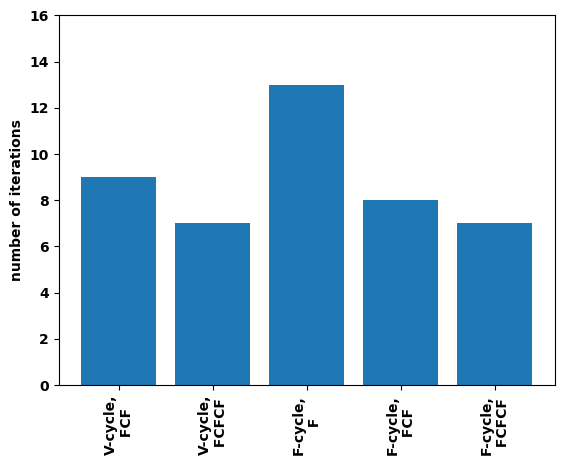}}
	\caption{Results of five MGRIT variants in terms of runtime (left) and number of iterations (right). The runtime results are shown for using up to 64 processes for parallelizing only in time.}
	\label{fig:example_1_results}
\end{figure}

Figure \ref{fig:example_1_results} shows the runtimes and number of iterations of the \texttt{PyMGRIT} simulations. The left figure presents the runtime as a function of the number of processes for up to 64 processes and the right figure shows the number of iterations required to reach the halting tolerance for the different MGRIT variants. Comparing only the number of iterations, we see that iterations decrease when stronger (and, thus, more expensive) relaxation schemes are used. Similarly, an $F$-cycle typically requires fewer iterations than a $V$-cycle with the same relaxation scheme due to the extra work in each iteration. As an example, the number of iterations for the $F$-cycle with $F$-relaxation is about one and a half times the number of iterations for with $F$-cycle variant with $FCF$-relaxation. Using the even stronger $FCFCF$-relaxation, the number can be reduced further, but at a smaller scale.

However, the choice of the relaxation scheme and the cycle type affects the cost per iteration. Looking at the runtimes of $V$-cycles, the runtime of the variant with $FCFCF$-relaxation is always higher than that of the variant with $FCF$-relaxation, although fewer iterations are required for the stronger relaxation scheme. Looking at $F$-cycles, we see that runtimes are higher than those of $V$-cycles for the same relaxation scheme, although iteration counts are smaller. Additionally, $F$-cycles scale worse than $V$-cycles.

\begin{figure}
	\centerline{\includegraphics[width=.32\linewidth]{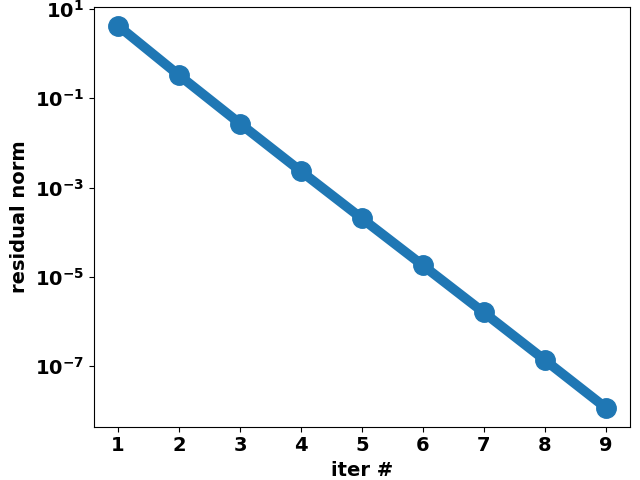}\quad
	\includegraphics[width=.32\linewidth]{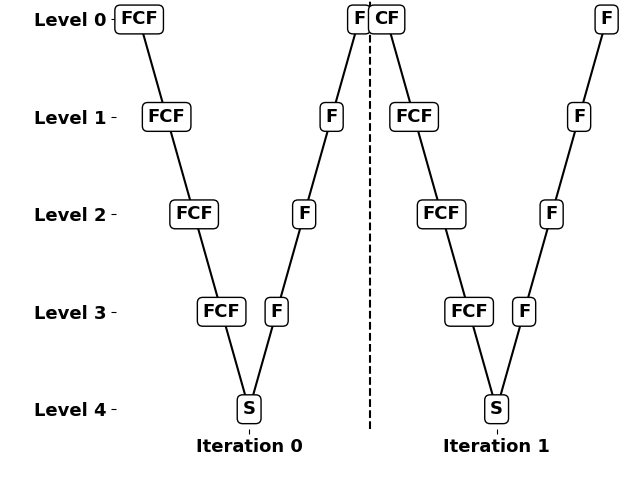}\quad
	\includegraphics[width=.32\linewidth]{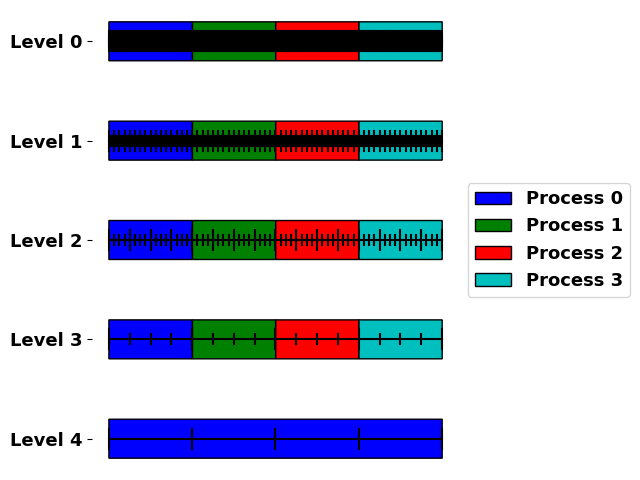}}
	\caption{Plots provided by \texttt{PyMGRIT}'s class \texttt{MgritWithPlots} for solving the one-dimensional heat equation using five-level MGRIT $V$-cycles with $FCF$-relaxation and using four processes in time. While the convergence plot (left) requires running a full \texttt{PyMGRIT} simulation, the plots visualizing the cycling strategy (middle) and the distribution of time points across processes (right) can be created directly after setting up the MGRIT solver.}
	\label{fig:pymgrit_statistic}
\end{figure}

Figure \ref{fig:pymgrit_statistic} shows three sample plots that can be automatically generated by the \texttt{PyMGRIT} class \texttt{MgritWithPlots}. Shown are the plots for the $V$-cycle variant with $FCF$-relaxation and using four processes in time. The diagram on the left shows the residual norm after each MGRIT iteration of a \texttt{PyMGRIT} simulation. The other two plots in Figure \ref{fig:pymgrit_statistic} can be created immediately after initializing the \texttt{MgritWithPlots} object. The middle plot visualizes the MGRIT cycle and the right plot presents the distribution of time points across processes for all MGRIT levels.

The figures demonstrate how easy it is to visualize different aspects of \texttt{PyMGRIT}'s MGRIT algorithm. More examples of how \texttt{PyMGRIT} can be used to create diagrams, e.\,g., plotting the (space-)time solution of a problem, can be found in the documentation.

\subsection{Time parallelism with PyMGRIT and GetDP}
\label{sec:time_parallelism_with_pymgrit_and_getdp}

In this example, we apply \texttt{PyMGRIT}'s MGRIT algorithm to a simulation of a two-dimensional electrical machine, whereby the solution of the spatial problem at each time step is executed by the external \texttt{GetDP} solver. The example demonstrates how an existing time integration routine from a complex simulation code can be integrated into the parallel \texttt{PyMGRIT} framework, enabling a dramatic reduction of the simulation time. This section summarizes the results presented in \cite{bolten2019parallelintime}; more details about the simulation can be found in \cite{bolten2019parallelintime} and all files are available in \texttt{PyMGRIT}'s folder \texttt{examples/induction\_machine}.

The standard approach for the simulation of electrical machines is given by the so-called eddy current problem, a simplification of Maxwell's equations in which the displacement current is neglected, and is defined in terms of the unknown magnetic vector potential $\mathbf{A}: \Omega \times \timeint \rightarrow \mathbb{R}^3$ as 
\begin{subequations}\label{eq:system_eddy_current}
\begin{alignat}{2}
		\sigma \partial_t \mathbf{A} + \nabla \times \left(\nu(\cdot) \nabla\times\mathbf{A}\right)&= \mathbf{J}_{s} &&\text{in} \; \Omega \times \timeint, \label{eq:system_eddy_current_a} \\
		\mathbf{n} \times \mathbf{A} &= 0 &&\text{on} \; \partial \Omega, \label{eq:system_eddy_current_b} \\
		\mathbf{A}\left(\mathbf{x},t_0\right) &= \mathbf{A}_0(\mathbf{x}), \label{eq:system_eddy_current_c}
\end{alignat}
\end{subequations}
with spatial domain $\Omega$, consisting of rotor, stator, and the air gap in between, time interval $\timeint$ and Dirichlet boundary conditions. The geometry is encoded by the scalar electrical conductivity $\sigma(\mathbf{x}) \geq 0$ and the (nonlinear) magnetic reluctivity $\nu(\mathbf{x},|\nabla\times\mathbf{A}|) > 0$. Three ($n_s = 3$) homogeneously distributed stranded conductors \cite{schoeps2013windingfunctions} are modeled by the source current density
\[
		\mathbf{J}_{s} = \sum_{s=1}^{n_{s}} \mathbf{\chi}_s i_s,
\]
with winding functions $\mathbf {\chi}_s: \Omega \rightarrow \mathbb{R}^3$ and currents $i_s: \timeint \rightarrow \mathbb{R}^3$. An attached electrical network provides a connection between the voltage $v_s\left(t\right), s=1,2,3$, and the so-called flux-linkage.

To consider the rotation of the rotor, the problem is extended by an additional equation of motion,
\begin{subequations}\label{eq:system_motion}
	\begin{alignat}{2}
	\omega(t)&= \frac{d\theta\left(t\right)}{dt}, \quad &&t \in \timeint, \label{eq:system_motion_a}\\
		I\frac{d^2\theta}{dt^2}+C\frac{d\theta}{dt}+\kappa \theta &= T_{\text{mag}}\left(\mathbf{A}\right) \quad &&\text{in} \; \timeint, \label{eq:system_motion_b} \\ 
		\theta\left(t_0\right)&=\theta_0, \label{eq:system_motion_c} \\
		\omega(t_0) &= \omega_0, \label{eq:system_motion_d}
	\end{alignat}
\end{subequations}
where the torque $T_\text{mag}$ defines the mechanical excitation, $\omega$ is the angular velocity of the rotor, $I$ denotes the moment of inertia, and $C$ and $\kappa$ are the fricition and torsion coefficients, respectively. The moving band approach \cite{FerreiraDaLuz_etal_moving_band} is used to model the movement of the mesh.

We reduce the three-dimensional ($3$D) domain $\Omega$ into a two-dimensional ($2$D) domain $\Omega_{2D} \subset \mathbb{R}^2$ in the $x,y$-plane and discretize the problem \eqref{eq:system_eddy_current}, combined with \eqref{eq:system_motion}, in space using linear finite elements with $n_a$ degrees of freedom. This yields a system of equations of the form
\begin{subequations}\label{eq:semi_discrete_system}
	\begin{align}
		M \mathbf{u}'(t) + K(\mathbf{u}(t)) \mathbf{u}(t) &= \mathbf{f}(t), \quad t \in \timeint \label{eq:semi_discrete_system_a}\\
		\mathbf{u}(t_0) &=\mathbf{u}_0, \label{eq:semi_discrete_system_b}
	\end{align}
\end{subequations}
with unknown $\mathbf{u}^\top=[\mathbf{a}^\top, \mathbf{i}^\top, \theta, \omega] : \timeint \rightarrow \mathbb{R}^n$. At one point in time $t \in \timeint$, the solution $\mathbf{u}\left(t\right) \in \mathbb{R}^n$ consists of the magnetic vector potential $\mathbf{a}\left(t\right) \in \mathbb{R}^{n_a}$, the currents of the three phases $\mathbf{i}\left(t\right) \in \mathbb{R}^3$, the rotor angle $\theta\left(t\right) \in \mathbb{R}$, and the angular velocity of the rotor $\omega\left(t\right) \in \mathbb{R}$. Note that problem \eqref{eq:semi_discrete_system}, due to the presence of non-conducting materials, consists of differential-algebraic equations (DAEs) of index-1 \cite{Bartel_2011aa,Cortes-Garcia_2019aa}.


We use the multi-slice finite element model ``im\_$3$\_kw'' \cite{JGyselinck2001_multi_slice} of an induction machine for modeling the semi-discrete problem \eqref{eq:semi_discrete_system}. More precisely, a modified version of the machine \cite{Gander_2019aa} supplied by a three-phase pulse width modulated voltage source of 20 kHz is considered. We refer to \cite{bolten2019parallelintime} for more details. Gmsh \cite{gmsh, gmsh_website} is used to construct the mesh representation of the model and a hierarchy of two nested meshes is considered to allow for spatial coarsening. The fine mesh $\mesh_1$, consisting of $n_a=17{,}496$ degrees of freedom, is constructed by refining the coarser mesh $\mesh_2$ with $n_a=4449$ degrees of freedom. The time step routine calls the \texttt{GetDP} solver, which implements the time integration using backward Euler.

We apply five different MGRIT variants to the simulation, choosing $\mesh_1$ as the spatial grid, a time step $\Delta t = 2^{-20}$ and $N_t=10{,}753$ time steps, resulting in a final time $t_f \approx 0.01$ s. For all MGRIT variants, we choose a convergence criterion based on the relative change of joule losses at $C$-points of two consecutive iterations. The algorithm stops if the maximum norm of the relative difference of two successive iterations is less than 1\%. The following MGRIT variants are chosen: a two-level MGRIT $V$-cycle with $F$-relaxation, a five-level MGRIT $V$-cycle with $FCF$-relaxation and a five-level MGRIT $F$-cycle with $FCF$-relaxation, whereby both five-level variants are applied with and without spatial coarsening. Further, a non-uniform temporal coarsening strategy with coarsening factor 42 on the first level and, for the multilevel variants, a coarsening factor of four on all coarse levels is chosen.

\begin{figure*}
	\begin{center}
	\includegraphics[width=0.7\textwidth]{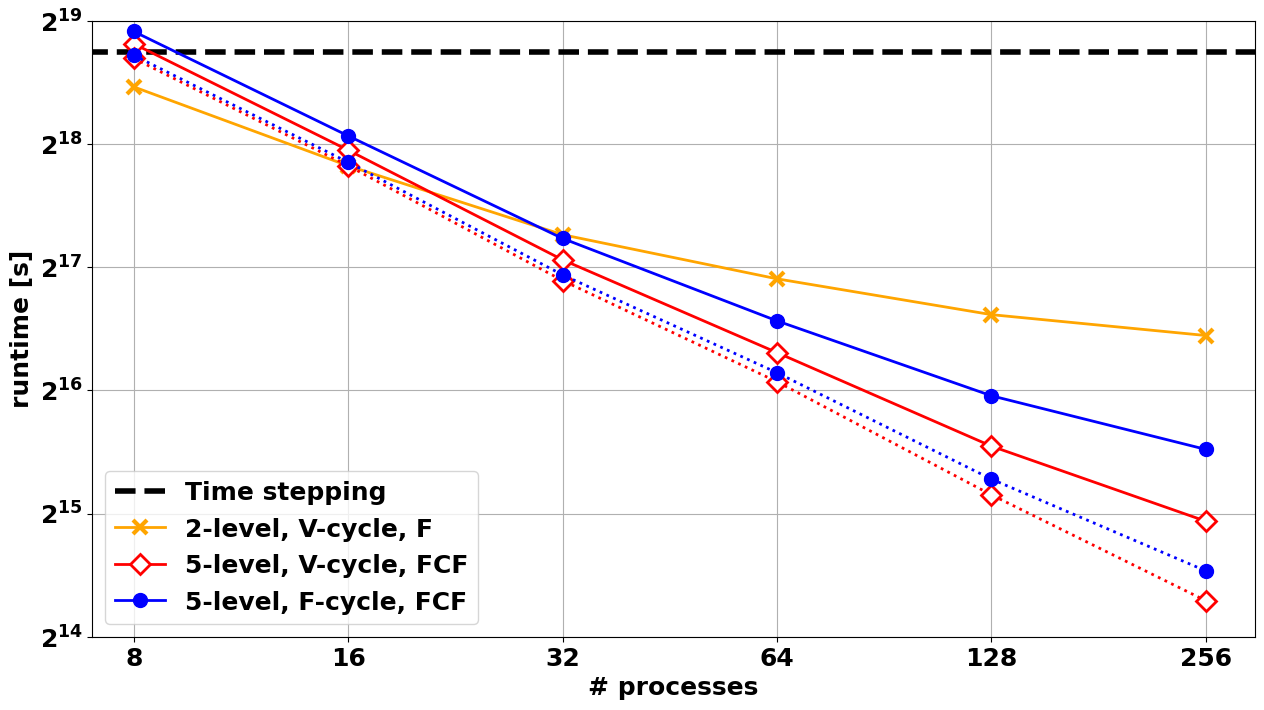}
		\caption{Runtimes of five different MGRIT variants and sequential time stepping for the nonlinear electrical machine ``im\_$3$\_kw''. Solid lines represent variants without spatial coarsening and dotted lines with spatial coarsening. For reference purposes, the dashed line shows the runtime for sequential time stepping on one processor.}
	\label{fig:example_2_strong_scaling}
	\end{center}
\end{figure*}

Figure \ref{fig:example_2_strong_scaling} shows the runtimes for the five different MGRIT variants as a function of the number of processes and, for reference purposes, the runtime for sequential time stepping on one processor. Thereby, the dashed line represents the runtime results for sequential time stepping, which is about five days, solid lines represent the MGRIT variants without spatial coarsening and dotted lines with spatial coarsening. While the multilevel variants require about eight processes to achieve the same runtime results as the sequential method, the two-level variant with eight-way parallelism achieves a runtime of about four days. However, the multilevel variants show better strong scaling, such that the multilevel variants and the two-level variant using 16 processes have approximately the same runtime. Increasing the number of processes to 256, the runtime of the fastest multilevel variant is about 5.5 hours, which corresponds to a speedup of about 22 compared to sequential time stepping.


\subsection{Space-time parallelism with PyMGRIT and PETSc}
\label{sec:space_time_parallelism_with_pymgrit_and_petsc}

In the two previous examples, only parallelization in time was considered. In the last example, we show results for space-time parallel runs with \texttt{PyMGRIT}. There are several approaches for extending \texttt{PyMGRIT} with spatial parallelism, e.\,g., the use of external libraries that take care of spatial parallelism. Two examples, the coupling of \texttt{PyMGRIT} with \texttt{Firedrake} \cite{Firedrake} and with \texttt{PETSc} \cite{petsc-user-ref}, are available in the online documentation and in the repository. In this example, we consider the coupling of \texttt{PyMGRIT} with \texttt{PETSc} via the Python package \texttt{petsc4py} \cite{DALCIN20111124}, which allows access to \texttt{PETSc} data types, solvers and MPI-based parallization in space without having to use lower-level programming languages like Fortran or C\texttt{++}. Note that \texttt{petsc4py} is not automatically installed with the installation of \texttt{PyMGRIT}, but can be easily installed using \texttt{pip}. The code for the example can be found in the \texttt{examples/toms} directory.

To demonstrate space-time parallelization we choose a standard example of parallel-in-time integration methods \cite{Gander2015_Review}. More precisely, we choose the the forced 2D heat equation,
\[
    u_t - \Delta u = b(x,y,t) \;\; \text{ in } \; [0,1]\times[0,1]\times(t_0,t_{f}],
\]
with initial condition  $u(x,y, t_0) = u_0(x,y)$, homogeneous Dirichlet boundary conditions, and forcing term $b(x,y,t)$ such that the exact solution is given by
\[
    u(x,y,t)=\sin(2\pi x)\sin(2\pi y)\cos(t) \;\; \text{ in } \; [0,1]\times[0,1]\times[t_0,t_{f}].
\]

The problem is discretized using standard central finite differences in space with $N_x=129^2$ degrees of freedom. In time, the problem is discretized on a grid with $N_t = 16{,}385$ points using backward Euler.

We apply two five-level MGRIT variants to the problem, a $V$-cycle with $FCF$-relaxation and an $F$-cycle with $F$-relaxation. Both variants use a coarsening strategy with varying coarsening factors $m_1=32$, $m_2=16$, $m_3=4$, $m_4=4$, i.\,e., coarsening between the finest and the first coarse level is applied using a factor of $32$, then a factor of $16$ is used, and so on. The stopping criterion for both variants is based on the discrete 2-norm of the space-time residual with a tolerance of $10^{-7}$. For both variants, the nested iteration strategy is used to get an improved initial guess.

For the implementation of the \texttt{PyMGRIT} application and vector classes for the 2D heat equation, \texttt{PETSc} structures can be used thanks to the \texttt{PETSc} backend, such that the vector class uses DMDA global vectors as data structure to store the solution at a point in time and the time-stepping method in the application class uses the linear Krylov space (KSP) solver of \texttt{PETSc}. Then, space-time parallelism is easily achieved by partitioning the \texttt{MPI\_COMM\_WORLD} communicator into two communicators: one time and one space communicator.

\begin{figure*}
	\begin{center}
	\includegraphics[width=0.75\textwidth]{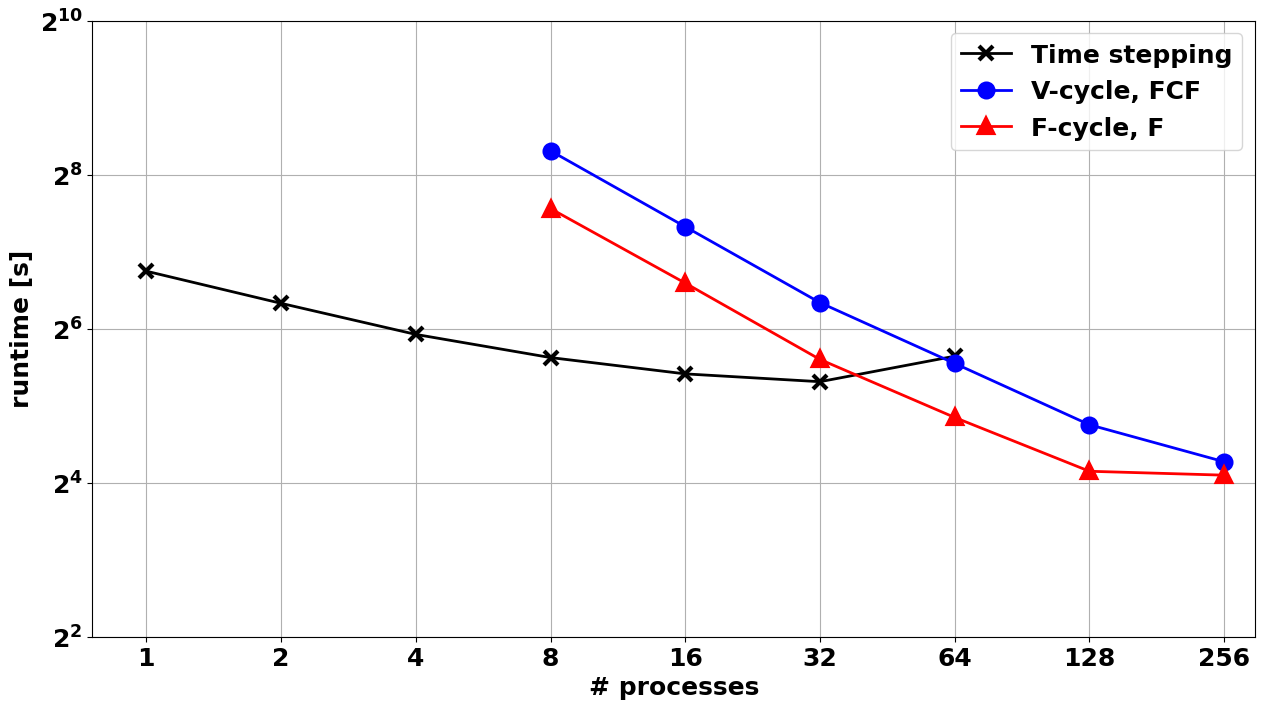}
		\caption{Strong scaling results for two five-level MGRIT variants and space-parallel time-stepping. For the multi-level variants, four processes in space are used, i.\,e., 16 processes correspond to four processes in time and four processes in space.}
	\label{fig:example_3_space_time}
	\end{center}
\end{figure*}

Figure \ref{fig:example_3_space_time} presents strong scaling results for the two MGRIT variants and additionally for a time stepping algorithm, with parallelization only in space. The time-stepping method achieves the fastest runtime for 32 processes in space, so this would be the optimal value to add time parallelism. However, due to the limited number of cores on the cluster, we select only four processes in space for the two MGRIT variants, such that 64 processes in the plot correspond to using four processes for spatial parallelization and 16 processes in time for the MGRIT variants, i.\,e., a total of $4*16=64$ processes. Note, that a larger number of processes in space could bring further improvement.  With a small number of processes, time stepping is faster than the multi-level variants, but the MGRIT variants show significantly better results with increasing numbers of processes, whereby the crossover point is reached at approximately 64 processes. For 256 processes the speedup of the fastest MGRIT variant is about three compared to space-parallel time-stepping with 32 processes. This could be improved by using even more processes in space and time. 

\section{Conclusion and outlook}
\label{sec:conclusion_and_outlook}

This paper introduces the Python framework \texttt{PyMGRIT}, which implements the parallel-in-time method multigrid-reduction-in-time (MGRIT). The \texttt{PyMGRIT} framework allows the application of the MGRIT algorithm without having to worry about implementation details, parallel communication and so forth, as well as easy prototyping of new variants of the algorithm, and (space-)time parallel simulations using MPI. In addition to the Python code, whose functionality is guaranteed by a continuous integration environment with automated serial and parallel tests, the framework also offers extensive documentation with a quickstart code example, a tutorial, and many examples demonstrating various features of \texttt{PyMGRIT}. Together with a simple installation and many pre-implemented ordinary and partial differential equations, the documentation provides an easy start into working with the package. Advanced usage is described in a separate section of the documentation, and more information about \texttt{PyMGRIT}'s core classes and functions can be found in the API documentation, which is automatically generated from Python comments. By coupling \texttt{PyMGRIT} with other libraries, such as \texttt{Firedrake} or \texttt{PETSc}, the time parallelism provided by the package can be combined with spatial parallelism for space-time parallel simulations.

The \texttt{PyMGRIT} package already offers many features; however, there are still many possible extensions, open tasks, and new directions. In the following we summarize possible future work. Due to its structure and iterative nature, the MGRIT algorithm allows for a large number of variations and adaptations, e.\,g., considering different cycle types, applying different relaxation strategies on different MGRIT levels or for different MGRIT iterations, and so forth. While many of these MGRIT variants are already implemented in the \texttt{PyMGRIT} package, there are still some extensions, such as adaptive time stepping, that are not implemented yet. Furthermore, for nonlinear problems, options such as the additional storage of an auxiliary vector \cite{MR3716560} that provides an improved starting solution for nonlinear solvers, are attractive extensions.

Another area is the improvement of the required storage space. The MGRIT algorithm, as implemented in \texttt{PyMGRIT} requires more memory than sequential time stepping, because the solution is stored for multiple points in time. There are different strategies to reduce the memory costs, e.\,g., storing the solution only at $C$-points, but these are not implemented yet and are topics for future work.

\texttt{PyMGRIT} contains examples of coupling the package with \texttt{Firedrake}, \texttt{PETSc} and \texttt{GetDP}, but of course many other powerful libraries exist. Coupling with more libraries could increase the number of applications for the package and make it easier for users to implement their specific problem. We mention here the libraries \texttt{FENiCS} \cite{LoggMardalEtAl2012a} and \texttt{PyClaw} \cite{pyclaw-sisc}, which both provide a wide range of space-specific algorithms.

\begin{acks}
The authors are grateful to Jacob Schroder and Robert Falgout for many helpful conversations and insightful comments regarding the implementation of the MGRIT algorithm, and would like to thank Ryo Yoda for sharing his experiences of using \texttt{PyMGRIT} that led to improvements of the package. The authors also acknowledge the support by the BMBF in the framework of project PASIROM (grant number 05M18PXB).
\end{acks}

\bibliographystyle{ACM-Reference-Format}
\bibliography{pymgrit}

\end{document}